\documentclass[11pt]{iopart}  
\pdfoutput=1
\usepackage{graphicx}
\usepackage{upgreek}
\usepackage[margin=2.5cm]{geometry}
\begin{document}
\article[A new dimension for piezo actuators]{FAST TRACK COMMUNICATION}{A new dimension for piezo actuators: Free-form out-of-plane displacement of single piezo layers}
\author{Matthias C Wapler, Jens Brunne and Ulrike Wallrabe}
\address{Laboratory for Microactuators, Department of Microsystems Engineering \\
University of Freiburg, 79110 Freiburg, Germany}
\ead{e-mail: wallrabe@imtek.uni-freiburg.de}
\begin{abstract}We present a controlled mode of ``topological'' displacement of homogeneous piezo films that arises solely from an inhomogeneous in-plane strain due to an inhomogeneous polarization. For the rotationally symmetric case, we develop a theoretical model that analytically relates the shape of the displacement to the polarization  for the cases of in-plane and  out-of-plane polarization. This is verified for several examples, and we further demonstrate controlled asymmetric deformations.
\end{abstract}
% Keywords: PZT films, bending actuators, free-form displacement, out-of-plane displacement}
\submitto{\SMS}
% \maketitle
% \abstract{We describe }
\section{Introduction} 
Piezoelectric materials have been widely used since the middle of the $20^{th}$ century in a myriad of actuators in many applications ranging from micro-pumps (see e.g. \cite{micropumps}) to adaptive wavefront correction in astronomical telescopes (e.g. \cite{astronomy}). 
They are typically used directly through their longitudinal or transverse strain, via lever mechanisms for example in flextensional ``moonie'' actuators \cite{moonie}, or as
bending actuators where the displacement arises from a difference in the strains of different planes in the actuator. 
% The latter are either composites where piezo films are joined with passive layers or differently polarized piezo films --- for example metal laminates (``THUNDER'', \cite{thunderpat,thunderpaper}), laminates with passive fibers (``LIPCA'' \cite{lipca}) or active fiber materials \cite{piezofibers} --- or they are based on functional gradient materials (``RAINBOW'' \cite{rainbow}). 
The latter are usually composites where piezo films are joined with passive layers or differently polarized piezo films -- for example metal laminates  \cite{thunderpat,thunderpaper}, laminates with passive fibers  \cite{lipca} or active fiber materials \cite{piezofibers}. Another mechanism are  functional gradient materials  \cite{rainbow}.
In any event, the full potential of the possible strain configurations of the material is not used.

In this paper, we study how specific inhomogeneous polarizations can yield out-of-plane displacements of single, homogeneous piezoelectric films. In contrast to the above outlined bending principle, these displacements originate solely from the inhomogeneous in-plane strain in the ``intrinsic'' geometry of the piezo film, which causes its embedding in the ``extrinsic'' geometry to deflect. From an application point of view, this mode of displacement allows for actuators with free-form displacements that can have relatively sharp features with small bending radii compared to ordinary bending actuators. 
% From a topological point of view, these sharp features are actually the embedding artefacts of topological defects in the intrinsic geometry, for example a conical singularity yielding a conical tip.
% The displacement is also to leading order independent of the material thickness, in contrast to the inverse scaling in the bending-type actuators. 
From a topological point of view, these sharp features are actually the embedding artefacts of topological defects in the intrinsic geometry. For example a conical singularity yields a conical tip.
Furthermore, the displacement is -- to leading order -- independent of the material thickness, in contrast to the inverse scaling in the bending-type actuators. 
Certainly, this principle can be generalized to other strain-type actuations of films and thin sheets.

First, we develop a theoretical description of the displacement for rotationally symmetric configurations for in-plane polarized piezos, where both the longitudinal and the transverse indirect piezo effect contribute, and for polarizations orthogonal to the plane, where only the transverse effect contributes. Then, we show how to implement the desired polarization through an appropriate electrode layout and briefly describe our rapid prototyping method, the materials and the experimental setup. Finally, we produce piezo films with different radial displacement profiles and compare the observed displacements to the theoretical predictions; and as an outlook we show how to implement a rotationally asymmetric pyramid-like structure.

\section{Theory}\label{theory}
The working principle is to create globally inhomogeneous and aniotropic strains in the piezo sheets, that cause the sheets to deform out of plane. In the following, we will in particular consider  rotationally symmetric configurations where the piezo expands in the radially and contracts tangentially (or at least expands less than in the radial direction).

Let us assume that we have a rotationally symmetric sheet that is strained in the radial and tangential directions,
\numparts\begin{eqnarray}
dr \ \rightarrow \ dr' & = & dr + \delta dr \ = \ dr (1+\varepsilon_r)\ \ \mathrm{and}  \\
dl \ \rightarrow \ dl' & = & dl + \delta dl \ = \ dl (1+\varepsilon_l)\ , 
\end{eqnarray}\endnumparts
with a corresponding circumference and radial distance
\numparts\begin{eqnarray}
l'(r)& = & \oint_{r=const}\!\!\!\!\! dl' \  = \ l(r) + \delta l(r)\  = \ l(r) (1+\varepsilon_l(r)) \ \ \mathrm{and} \label{lscale} \\
r'(r) & = &  \int \! dr' \ = \ r(1+\frac{1}{r}\int_0^r \! \varepsilon_r( \tilde r) d \tilde r) \ . \label{rscale}
\end{eqnarray}\endnumparts
 Then, for an inhomogeneous deformation, we may have a deficit angle $\delta \theta =\frac{l'(r)}{r'(r)} - 2\pi \simeq \varepsilon_l(r') - \frac{1}{r}\int_0^r \varepsilon_r( \tilde{r}) d \tilde{r}$. A non-vanishing deficit angle represents an intrinsic curvature in the piezo film, which manifests itself in an extrinsic curvature of the piezo embedding. If $\lim_{r\rightarrow 0} \delta \theta(r)\neq 0 $, then there exists a conical singularity, resulting in a conical tip of the embedding.
 
\begin{figure}
\includegraphics[width=0.37\textwidth]{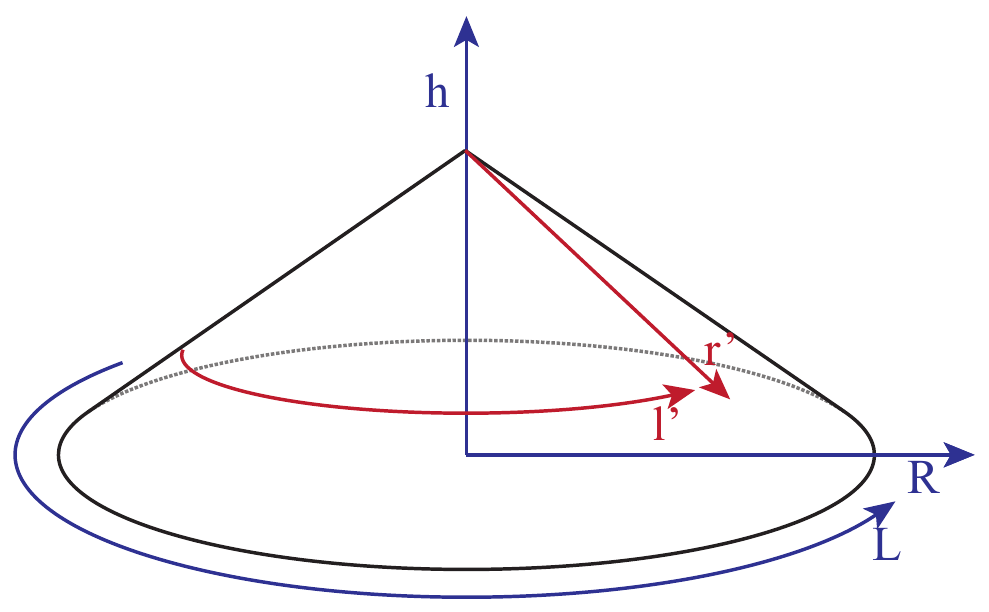}
\caption{Displacement of the piezo sheet with the intrinsic (red, $r', \, l'$) and extrinsic (blue, $h, \, R, \, L$) coordinates.}\label{thfig}
\end{figure}
To obtain the embedding in the extrinsic space as shown in fig. \ref{thfig}, $R(r)$ and $L(r)$ in the plane of the piezo and the displacement $h(r)$ orthogonal to it, we identify first of all $L(r) = l'(r)$, such that 
$R(r) = L(r)/(2\pi) = l'(r)/(2\pi)$, where we assumed small displacements. Then, we also notice that 
\begin{equation}
r'(r)=\int_0^r \sqrt{(\partial_{\tilde{r}} R(\tilde r))^2 + (\partial_{\tilde r} h(\tilde r))^2}\, d \tilde r \, \simeq \, R(r) \, + \, \frac{1}{2} \int_0^r \frac{(\partial_{\tilde r} h(\tilde r))^2}{\partial_{\tilde r} R(\tilde r)}\, d \tilde r \ .
\end{equation}
Substituting $R = l'/(2\pi)$, $l = 2 \pi r$ and equations \ref{lscale} and \ref{rscale}, and restricting ourselves to leading order in the deformation gives us
\begin{equation}
r(1+\frac{1}{r}\int_0^r \! \varepsilon_r( \tilde r) d \tilde r) \, = \, r (1+\varepsilon_l(r)) \, + \, \frac{1}{2} \int_0^r (\partial_{\tilde r} h(\tilde r))^2 \, d \tilde r \ ,
\end{equation}
which we differentiate to obtain
\numparts\begin{eqnarray}
1+ \varepsilon_r (r)  & \simeq &  1 + \partial_r (r\, \varepsilon_l(r))\, + \, \frac{1}{2} (\partial_{ r} h( r))^2 \ \ \mathrm{or} \\ \label{strainbend}
\partial_{ r} h( r) & \simeq & \pm \sqrt{2} \sqrt{\varepsilon_r (r) -\varepsilon_l (r) - r \partial_r \varepsilon_l (r)  } \ .
\end{eqnarray}\endnumparts
There, we see already that on the one hand, the direction of the displacement is (in the ideal case) arbitrary and on the other hand, not all intrinsic deformations yield real solutions so there are not always rotationally symmetric embeddings. 

Now, there are two most obvious ways to create these strains. Either we polarize the piezo sheet radially in-plane, with annular interdigitated electrodes, or we polarize it orthogonal to the plane, with planar ring-electrodes. 
% For the latter, we can create an annular interdigitated electrode structure and adjust the field strength, and thus the radial polarization, through the spacing of the fingers. 
For the former, the strain is for an initially un-polarized film with a subsequently applied electric field $E(r)$ given by $\varepsilon_r(r) = d_{33}|E(r)|$ and $\varepsilon_l(r) = d_{31}|E(r)|$, respectively. Hence equation \ref{strainbend} becomes
\begin{equation} \label{inplaneeq}
\left(\partial_{ r} h( r) \right)^2 \simeq 2 (d_{33} - d_{31})|E(r)| - 2 d_{31} r \,\partial_r |E(r)| \ ,
\end{equation}
which  simplifies further if we assume $d_{31} = -  d_{33}/2$. The resulting electric field distribution may be implemented by solving for $E(r)$ with an appropriate choice of the integration constant and then determining the radial positions $r_n$ of the electrodes through the potential difference between the electrodes $\Delta U \, = \, \int_{r_n}^{r_{n+1}} E(r) \, d r $, starting from some initial radius $r_0$. 

For the out-of-plane polarization with a transverse electric field, the strains are $\varepsilon_r(r) = d_{33}|E(r)| = \varepsilon_l(r)$, so the displacement is  given by 
\begin{equation} \label{outplaneeq}
\left(\partial_{ r} h( r) \right)^2 \simeq - 2 d_{31} r \, \partial_r |E(r)| \ ,
\end{equation}
which may be implemented by annularly structuring the electrodes on either side and using each $n$ and $m$ different potentials, such that one can obtain $n\times m$ different field strengths. The regions of the electrodes may be chosen by the appropriate mean or median field strengths or by the maximum deviation from the optimal field strength. 

This purely geometric derivation takes care of the in-plane forces for free boundary conditions but does not take into account bending moments and external forces. Pre-deflections and the effect of remanence and hysteresis may be taken into account for example by adjusting $d_{ij}$ and taking $E \rightarrow E + E_0$ in equations \ref{inplaneeq} and \ref{outplaneeq}, keeping in mind that the parameter $E_0$ depends on the field strength at which the polarization took place, or more generally on the history of the material.
For power-law profiles, the equations \ref{inplaneeq} and \ref{outplaneeq} and the appropriate radial positions of the electrodes have simple analytic solutions, but even in other cases the numerical solution is straightforward, so will not discuss these solutions further.

\section{Processing and Measurements}\label{process}
We will produce PZT prototypes with a thickness of $\sim 120\, \upmu \mathrm{m}$ and a diameter in the range of  $15 \, \mathrm{mm}$ to test these principles.

As a substrate, we use commercially available $ 120 \pm 20 \, \upmu \mathrm{m}$ thick PZT ceramic disks with a diameter of 25 mm and $ 5\, \upmu \mathrm{m}$ thick silver electrodes on both sides, that are commonly used -- laminated on metal sheets -- in acoustic transducers. These have on the one hand a low $\mathrm{T_c}$ of $\sim 250 ^\circ \mathrm{C}$ that is due to doping with $\sim 2.7\% $ strontium, enabling straightforward depolarization, and on the other hand a large coercive field strength of $\sim 950 \, \mathrm{V/mm}$. We further estimated a Young's modulus of $E \sim 50 \pm 8 \,  \mathrm{GPa}$ using a quick cantilever setup and a coefficient $d_{31} = -2.7 \pm 0.3 \times 10^{-4} \mathrm{mm/kV}$ based on a simple bending actuator.
% % 
% %  and a high $d_{31} = -2.7 \pm 0.3 \times 10^{-4} mm/kV$ coefficient that we determined from a simple bending actuator. This large $d_{31}$ is due to a special composition as the ceramic is doped with $\sim 2.7\% $ strontium \cite{} that replaces some of the lead content. It comes however at the expense of a reduced stiffness {\bf that we estimated in a simple cantilever experiment to $E \sim 50 \pm 8 GPa$. Eventually the latter properties are not of great relevance to this paper, as we simply compare the displacements relative to each other.

The fabrication takes place as follows: First, we structure the electrodes and then cut the contour by ablation with a UV laser, leaving contact pads with a size of approximately $2 \, \mathrm{mm}$.  Next, we remove the residues of the PZT and the silver in weak ultrasound in purified water with a dip of a few seconds in 25\% $\mathrm{HNO}_3$ at room temperature. The latter helps to break up the silver in the residues. For the in-plane polarized sheets, the PZT is then depolarized on a hotplate at nominally $420 ^\circ \mathrm{C}$, well-above the Curie temperature. For the out-of-plane polarization, we leave the initial polarization of the manufacturer.

\begin{figure}
\includegraphics[width=0.235\textwidth]{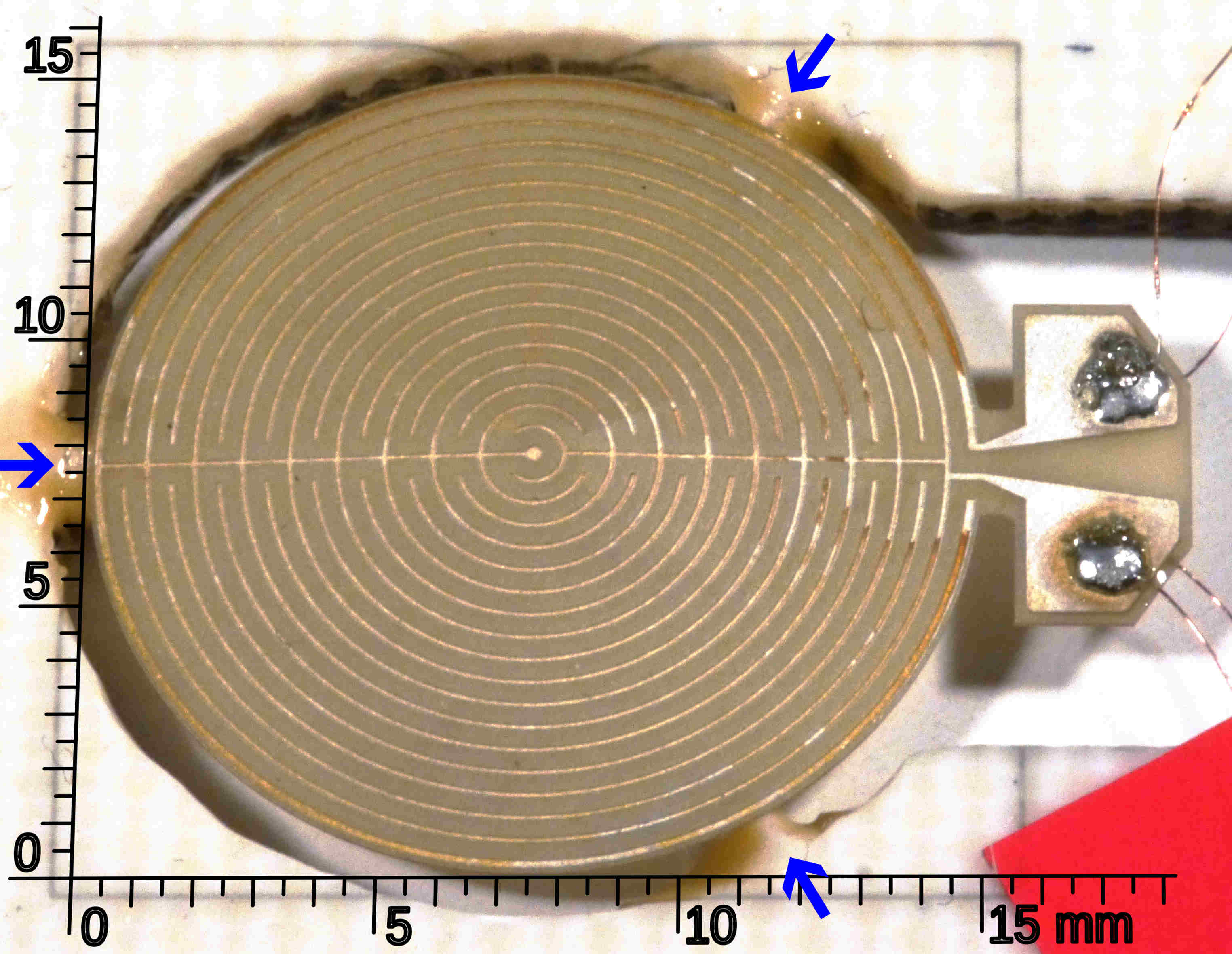}\hspace{0.02\textwidth}
\includegraphics[width=0.24\textwidth]{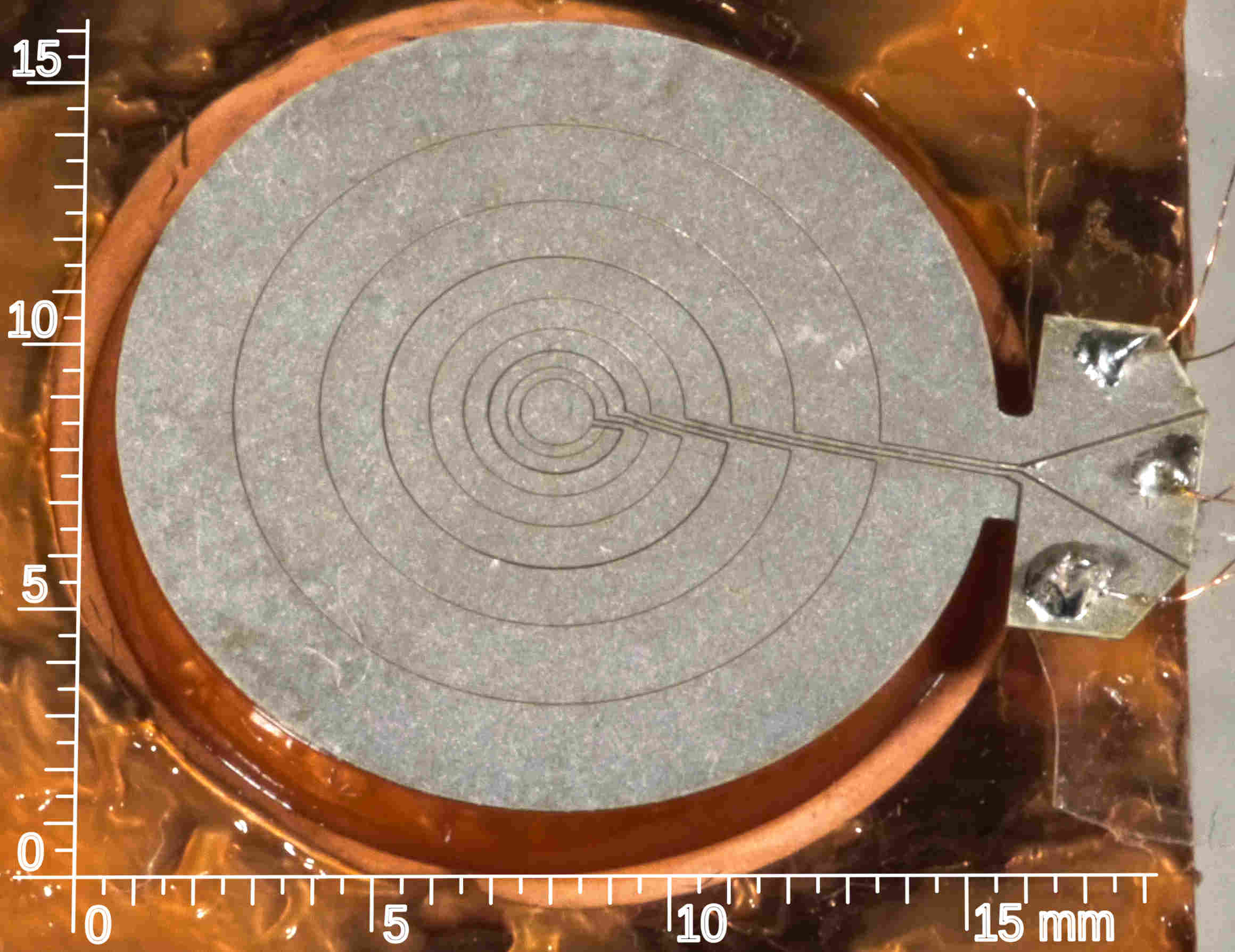}
\caption{In-plane polarized sheet on a 3-point mount (annular finger electrodes, mounting points indicated with arrows) and transversely polarized piezo on a supporting ring (circular planar electrodes, segmented with insulating grooves).}\label{imfig}
\end{figure}

For the characterization, the samples are supported at three points at their edge by adhesion on soft polyurethane with shore hardness A10 as shown in fig. \ref{imfig}. For some less stable configurations, they are supported by a $\sim 700\, \upmu \mathrm{m} $ wide and $\sim 1.5 \, \mathrm{mm} $ high supporting ring of the same material.
The characterization is then performed by scanning the surface of the piezo with a laser triangulation sensor at an applied DC voltage. The outer edge of the piezo is later  detected through the discontinuity in the measured profile and then gives us the center of the piezo and the correcponding radial profile.
\section{Experiments and Results}\label{characterization}
To test the expressions eq. (\ref{inplaneeq}) and (\ref{outplaneeq}) and explore this principle of deflection, we first investigate rotationally symmetric PZT films with in-plane polarization described by equation \ref{inplaneeq} for various different displacement profiles. At the end of the section, we also show a transversely (out-of-plane) polarized prototype described by eq. (\ref{outplaneeq}) and a demonstrator that goes beyond the assumed rotational symmetry.

\begin{figure}
\begin{center}\includegraphics[width=0.325\textwidth]{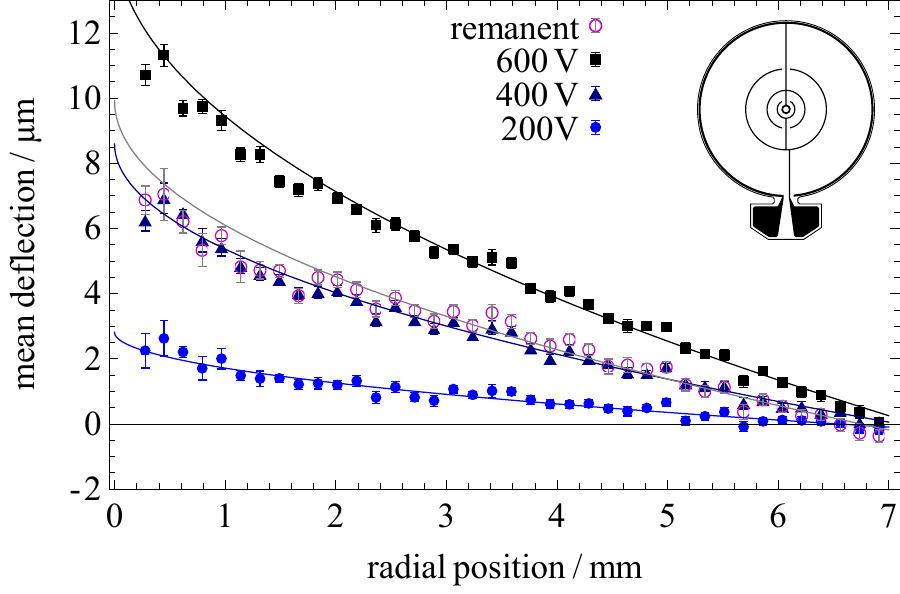}
\includegraphics[width=0.33\textwidth]{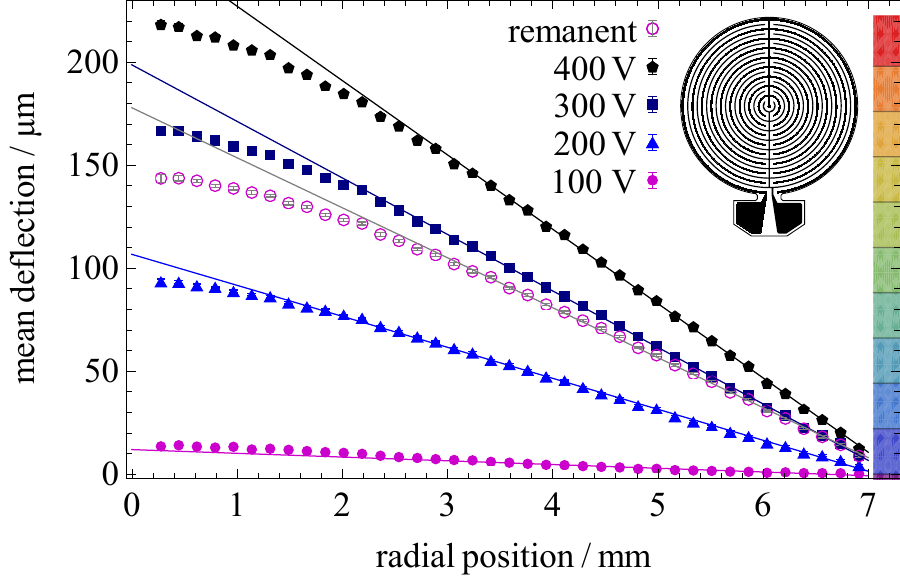}
\includegraphics[width=0.215\textwidth]{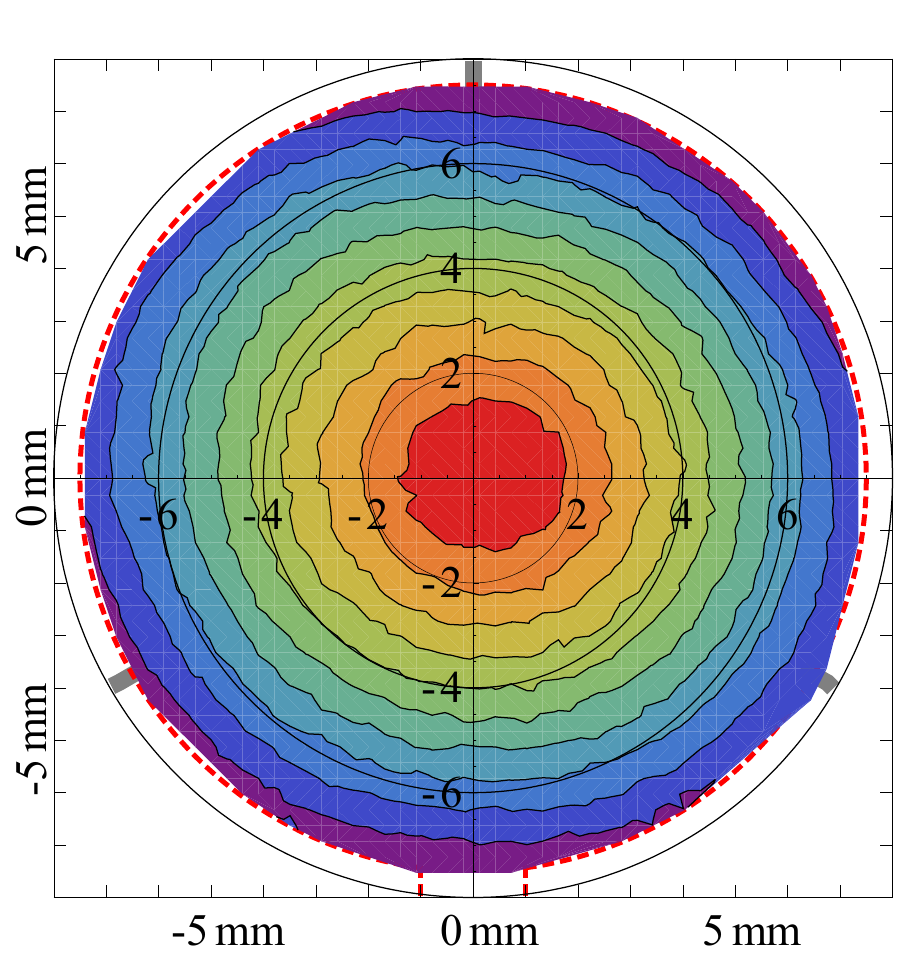}\\
\includegraphics[width=0.325\textwidth]{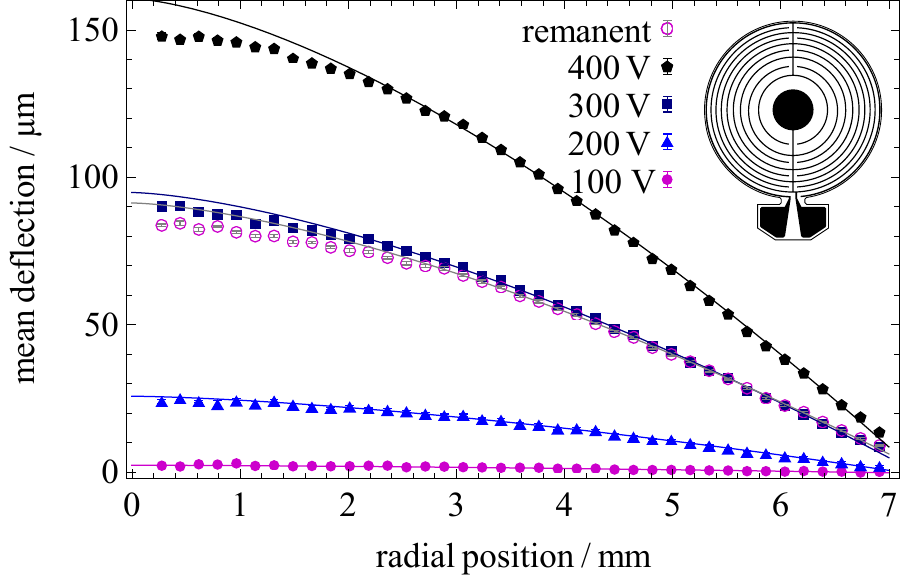}
\includegraphics[width=0.325\textwidth]{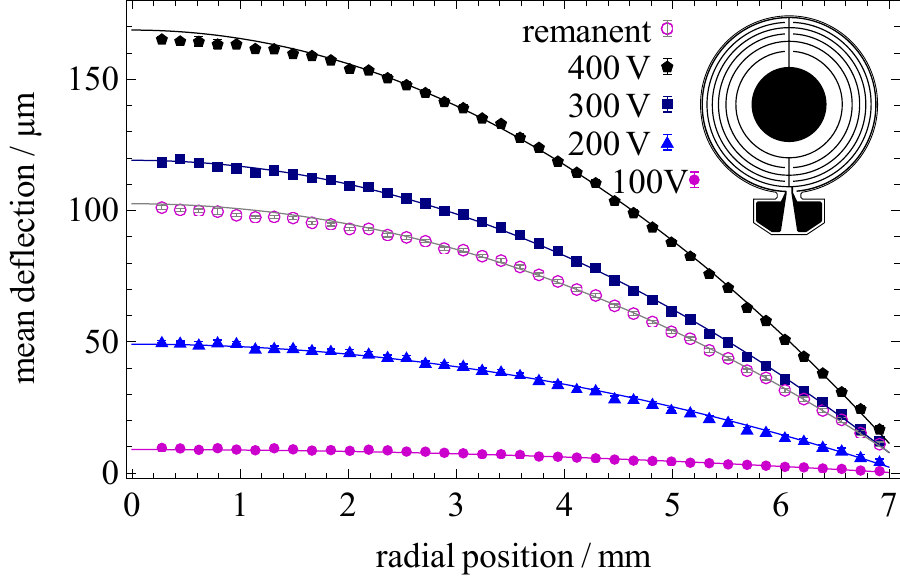}
\includegraphics[width=0.32\textwidth]{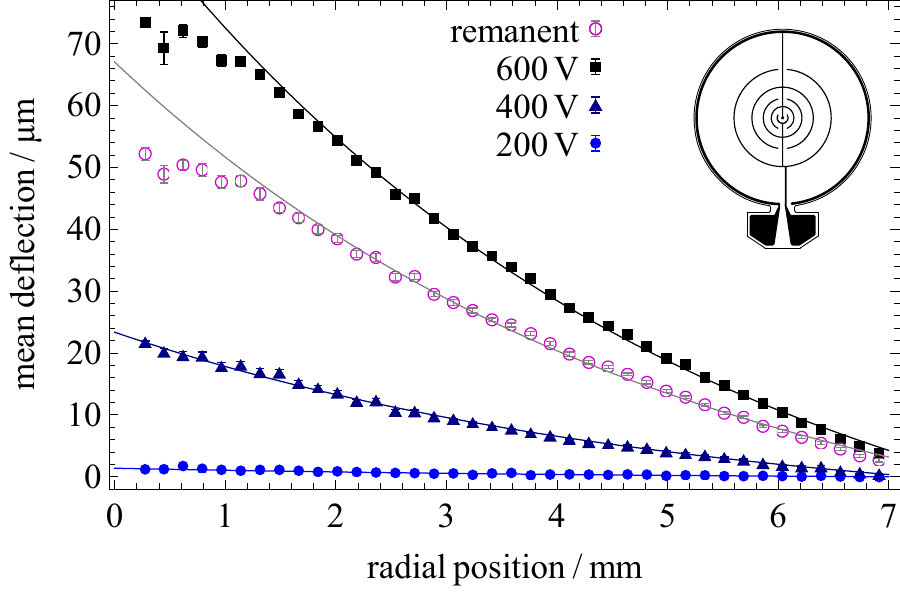}
\caption{Top left to bottom right: Radial profiles of piezo sheets with design profiles proportional to $r^{1/2}$, $r$, $r^{3/2}$, $r^2$, $e^{-r/5\mathrm{mm}}$ at various applied voltages, and the remanent displacement, after the highest shown voltage. The lines represent the scaled design profile and the insets show the piezo sheets (electrodes shown in black). The contour plot of the conical profile is fairly typical in symmetry; the red dashed line shows the perimeter of the piezo sheet and the gray bars represent the mounting spots.}\label{profiles}
\end{center}
\end{figure}

All of the prototypes have a diameter of approx. 15 mm. On the in-plane polarized films, we first structure single-sided electrodes with $80\, \upmu \mathrm{m}$ wide fingers as shown in the insets in fig. \ref{profiles}; the backside electrode was removed completely. The theoretical displacement profiles were the power laws $r^n$, $n\in \{1/2,1,3/2,2\}$ and the exponential $e^{-r/5 \mathrm{mm}}$. The linear profile which gives a conical displacement simply has a constant radial electric field strength, with electrode spacings of $320\, \upmu \mathrm{m}$. For the other profiles, we chose approximately $300\, \upmu \mathrm{m}$ as the narrowest electrode spacing. As the profiles with negative curvature tend to be unstable, i.e. asymmetric embeddings seem to be preferred, they were mounted on supporting rings. Also since the $\sqrt{r}$ profile has a singular derivative at the center and is hard 
to 
distinguish 
from a conical profile, a small hole was cut in the center 
of the piezo sheet. As we verified also for other displacement profiles, this reduces the overall displacement by around $30\%$, as there is missing material in the center to ``support'' the displacement.

Fig. \ref{profiles} shows the measured displacement profiles of initially un-polarized films at several applied voltages and the predicted profiles scaled to fit the overall displacement. We find that the displacements  fit the data very well, and the flattening in the central region increases with increasing displacements. In fact, fitting arbitrary power laws $r^\alpha$ or exponentials $r^{-r/\beta}$, respectively, to the outer region reproduces the predicted profiles very well with deviations in $\alpha$ less than $0.1$ for the power laws, depending on which part in the inner region is excluded; always below uncertainties in $\alpha$. The exponential yields a constant of $4\, \mathrm{mm}$ to $6.5\, \mathrm{mm}$, with higher values for higher voltages, which may be a result of saturation. The radial connecting electrodes and the material inhomogeneities may create a small deviation from rotational symmetry. This is shown for a typical example in the contour plot of a conical displacement in fig. \ref{profiles}, where we 
see only a small distortion at the position of the connecting electrodes.
\begin{figure}
\begin{center}\includegraphics[width=0.34\textwidth]{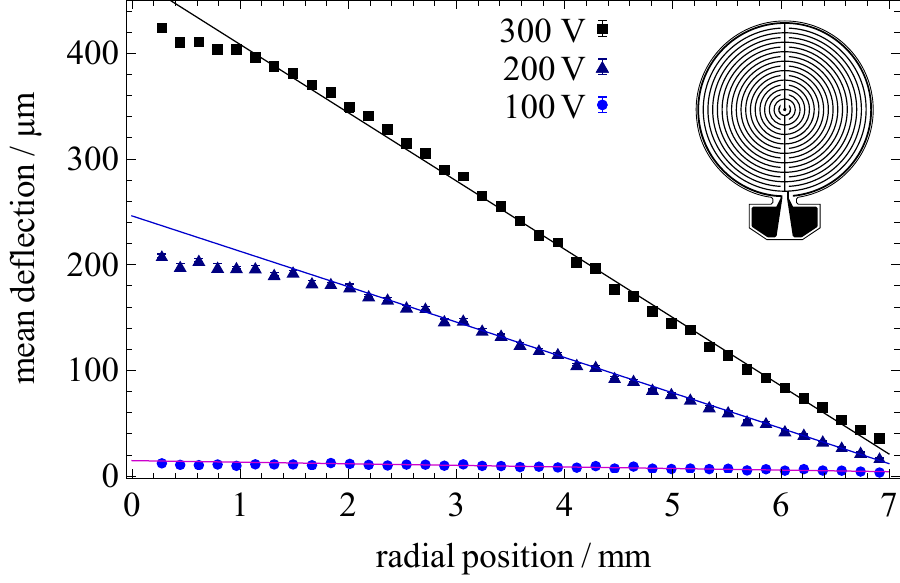}
\includegraphics[width=0.33\textwidth]{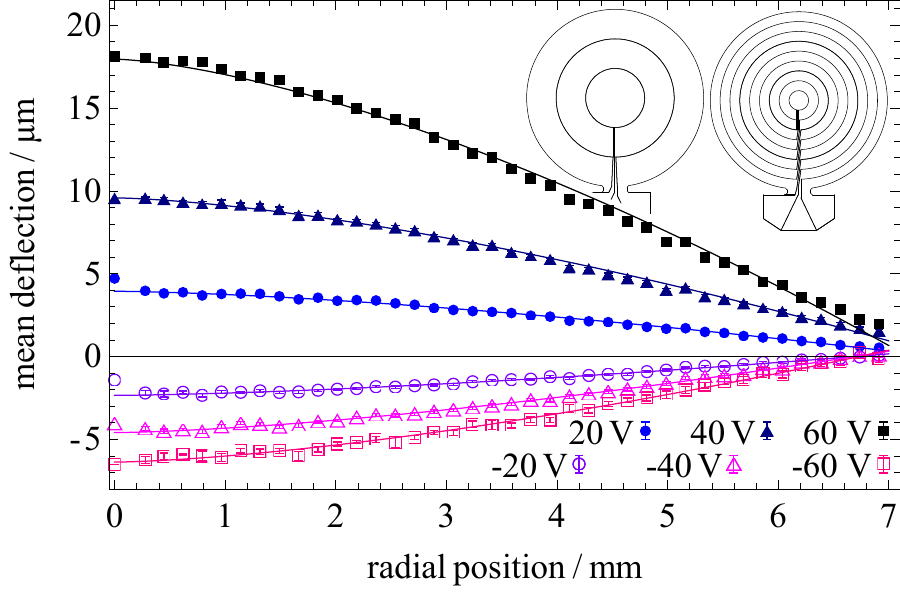}
\includegraphics[width=0.30\textwidth]{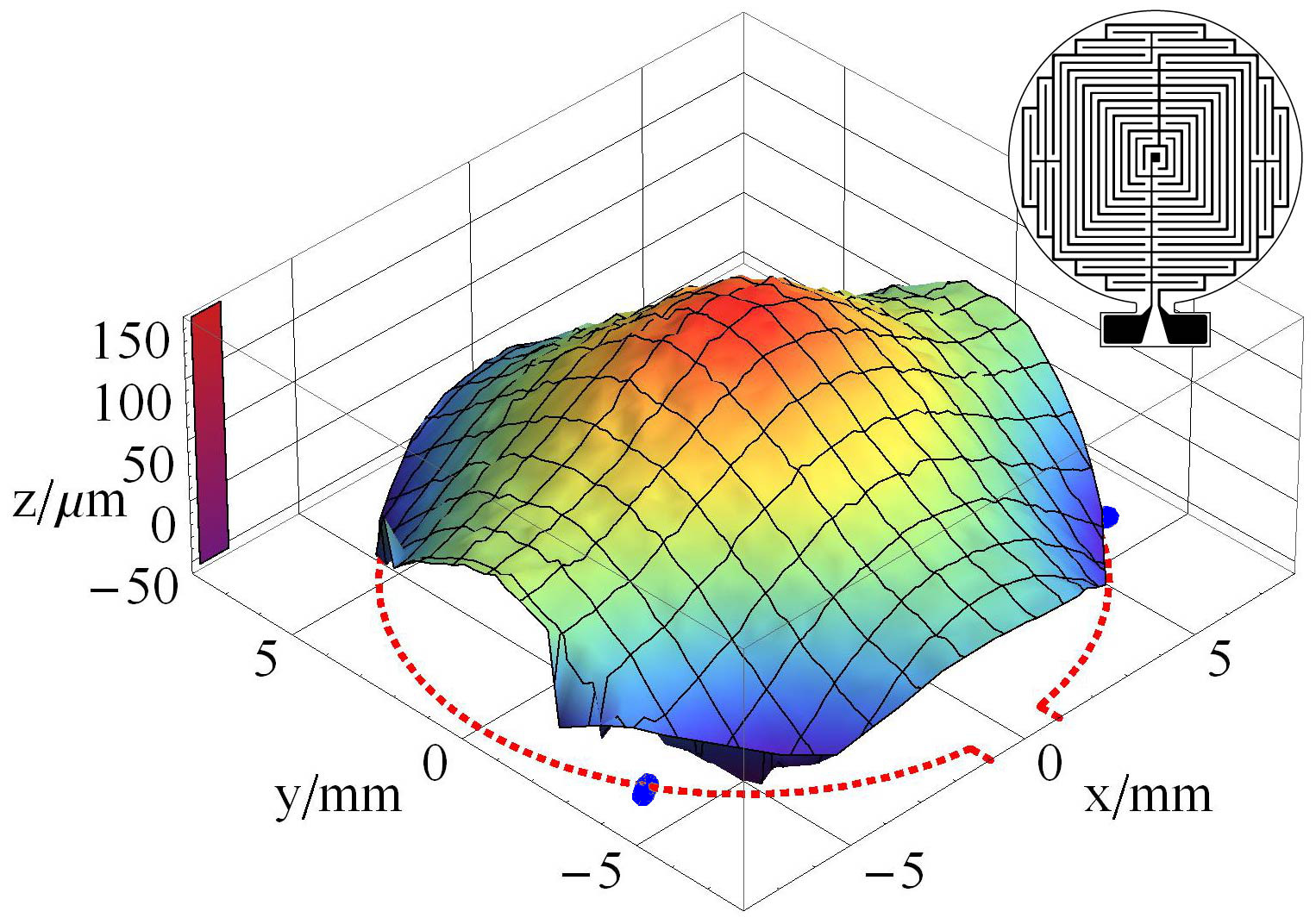}
\caption{Left to right: Conical radial profile originating from double-sided electrodes, radial profile of the out-of-plane polarized piezo sheet proportional to $r^{3/2}$ (lines in the inset represent the insulating grooves) and displacement due to a square-shaped electrode structure.}\label{moreprofiles}
\end{center}
\end{figure}

All the piezos were bent in the direction away from the electrodes, which may be an effect of a bending moment due to the asymmetry of the polarization of the single-sided electrodes. To exclude this bending moment as a source of the displacement, a piezo sheet designed for a conical displacement was structured with double-sided electrodes. The result is shown on the left in fig. \ref{moreprofiles} and displays an even smaller flattened region in the center. As seen already in the displacements of fig. \ref{profiles}, the overall displacement does not scale proportional to $\sqrt{E}$, in particular at small displacements. This arises probably from the fact that we have neglected the bending moments of the displaced piezo, nonlinearities and possible mechanical cracks. Still, the displacement at $200V$ corresponds by equation (\ref{inplaneeq}) to $d_{31}\sim -3.3\times 10^{-4} \mathrm{mm/kV}$, assuming an ideal radial polarization.

To test also the displacement of transversely polarized films described by eq. (\ref{outplaneeq}), a piezo sheet was structured with insulations of $30\, \upmu \mathrm{m}$ and $60\, \upmu \mathrm{m}$ in the electrodes to produce 3 electrodes on one side and 9 on the other side as shown in the inset in fig. \ref{moreprofiles}. As it is the most straightforward configuration, we chose a linear variation of the field strength, corresponding by eq. (\ref{outplaneeq}) to a displacement proportional to $r^{3/2}$. This was implemented through evenly spaced insulating grooves and potentials $\frac{1}{2}U_0,\ 0\ \mathrm{and} \  -\frac{1}{2}U_0$ on one side and $\frac{1}{6}U_0,\ 0\ \mathrm{and} \  -\frac{1}{6}U_0$ on the other side giving $-\frac{2}{3} U_0$ to $\frac{2}{3} U_0$ in steps of $\frac{1}{6} U_0$. We see that also in this case, the profile matches very well over most of the radial range. The flattening at the edge at $U_0 = 60V$ probably originates from the supporting ring. The overall displacement was 
significantly lower than 
predicted, as we could also see in the small 
displacements for the in-plane polarized mode.
As the film was initially homogeneously polarized, one can also apply a reverse field, resulting in a reversed strain. We see that this results in a negative relative displacement. This is because the film has a small initial displacement that was first straightened out by this strain, before the expected asymmetric displacements appear.

Finally, having discovered this new playground, we also produced a rotationally asymmetric configuration shown in fig. \ref{moreprofiles}. We see that the square-shaped electrode layout gives a pyramid-shaped displacement, rotated by $\pi/4$ as expected, since the steepest slope occurs where the divergence of the electric field is greatest. The overall displacement is similar to the cone in fig. \ref{profiles}.

\section{Summary and Conclusions}\label{conclusions}
We have successfully demonstrated controlled displacements of PZT sheets that were solely based on inhomogeneous intrinsic in-plane deformations of the sheet. For the rotationally symmetric case, we have derived expressions that describe the out-of-plane displacement resulting from radial in-plane polarizations and from out-of-plane polarizations. 
When we verified these expressions, we found that the profile of the displacements follows closely the predictions, up to some smoothing of regions with large curvatures or singularities. 

The overall amplitude of the displacement for large displacements is in the right order of magnitude but does not follow the predicted square-root scaling and particularly disagrees at small displacements. This is because our expressions are only based on the geometrical considerations of the embeddings, but ignore the bending moments. In particular at small field strengths, the displacements are large compared to the in-plane strains, such that the bending forces cannot be fully supported. Still, this does not affect the profiles: On the one hand the bending moments are local effects and the forces get distributed. From our discussion of the deficit angle and the conical singularity on the other hand, we see that this displacement mode is largely a global, topological, effect.

In principle, the direction of the displacement is arbitrary, but we have shown that with an asymmetric electrode layout, it is possible to control the direction through small ordinary bending effects. Further, we have demonstrated that it is possible to produce also displacements that are not rotationally symmetric.

In the future, we would like to study how our work can be generalized to generic displacements and how forces can be integrated in the model.
As our leading-order expressions are independent of the material thickness, it would be interesting at what curvature-thickness ratios the forces become relevant, and how the new thickness-independent scaling may play a role in microsystems.

Furthermore, it would be interesting to study whether this generic mode of displacement can also be implemented in other materials, e.g. electroactive polymers. 
It would be certainly also nice to see how this principle finds its way into applications, for example in micro systems or in new types of reflective adaptive optics.
\ack
The research of M.W. is financed by the Baden-W\"urttemberg Stiftung under the project ``ADOPT-TOMO'' and the research of J.B. is supported by DFG grant WA 1657/3-1.
% \begin{bibliography}
% 
% \end{bibliography}

\bibliography{piezobib}
\bibliographystyle{unsrt}
\end{document}